      [1999/12/01 v1.4c Il Nuovo Cimento]
\documentclass{cimento}

\usepackage{slashed}

\newcommand{\tr}{{\rm Tr}}
\newcommand{\st}{{\scriptscriptstyle T}}
\newcommand{\sL}{{\scriptscriptstyle L}}

\title{Universality of TMD distribution functions of definite rank}
\author{P.J.~Mulders\from{ins:1}
\thanks{Talk presented at Third Workshop on the QCD Structure of the Nucleon (QCD N'12), Bilbao, Spain, 22-26 October 2012}
\ETC,
M.G.A.~Buffing\from{ins:1}
\atque
A.~Mukherjee\from{ins:2}}
\instlist{\inst{ins:1}
Nikhef and Department of Physics, Faculty of Sciences,
VU University, Amsterdam, Netherlands
\inst{ins:2}
Department of Physics, Indian Institute of Technology, Bombay, Powai,
Mumbai 400076, India}
\PACSes{\PACSit{12.38.Aw, 12.38.-t, 13.88.+e}{}}

\begin{document}
\maketitle

\begin{abstract}
Transverse momentum dependent (TMD) distribution and fragmentation
functions are described as Fourier transforms of matrix elements
containing nonlocal combinations of quark and gluon fields.
These matrix elements also contain a gauge link operator with
a process dependent path, of which the process dependence that can be 
traced back to the color flow in the process. 
Expanding into irreducible tensors built from the transverse momenta 
$p_\st$, we can define a universal set of TMD correlators of definite 
rank with a well-defined operator structure.
\end{abstract}

\section{Introduction}

Transverse momentum dependent (TMD) distribution functions
(PDF) and fragmentation functions (PFF), also simply referred 
to as TMDs, basically are forward matrix elements of parton fields,
for instance for quarks
\begin{equation}
\Phi_{ij}(p|p) =
\int \frac{d^4\xi}{(2\pi)^4}\ e^{i\,p{\cdot} \xi}
\ \langle P\vert \overline\psi_j(0)\,\psi_i(\xi)\vert P\rangle ,
\label{phi-basic}
\end{equation}
including a summation over color indices.
For a single incoming fermion one would just have 
$\Phi \propto (\rlap{/}p + m)$. In a diagrammatic approach
in principle all sorts of correlators are needed, among them
quark-quark-gluon correlators defined
\begin{equation}
\Phi^\mu_{A\,ij}(p-p_1,p_1|p) = 
\int \frac{d^4\xi\,d^4\eta}{(2\pi)^8}
\ e^{i\,(p-p_1){\cdot} \xi}\ e^{i\,p_1{\cdot} \eta}
\ \langle P\vert \overline\psi_j(0)\,A^\mu(\eta)\,\psi_i(\xi)\vert P\rangle.
\label{quarkgluonquark}
\end{equation}
The basic idea is to factorize these hadronic (soft) parts in a full 
diagrammatic approach and parametrize them in terms of PDFs. This
will not work for the unintegrated correlators above. At high
energies, however, the hard scale provides for each hadron 
light-like vectors $P$ and $n$ such that $P{\cdot} n = 1$. For instance
$n = P^\prime/P{\cdot} P^\prime$, where $P^\prime$ is the momentum of another hadron obeying $P{\cdot} P^\prime \propto s$.
Using the light-like vectors, one writes down a Sudakov expansion 
of the parton momenta, $p = xP + p_\st + (p{\cdot} P - xM^2)n$
with $x = p^+ = p{\cdot} n$. In any contraction with vectors not 
appearing in the 
correlator, the component $xP$ contributes at order $\sqrt{s}$, the 
transverse
component at order $M$ and the remaining component contributes
at order $M^2/\sqrt{s}$.
This allows consecutive integration of the components. Starting
from the fully unintegrated result in Eq.~\ref{phi-basic} one
obtains the TMD light-front (LF) correlator 
\begin{equation}
\Phi_{ij}(x,p_\st;n) =
\left. \int \frac{d\xi{\cdot} P\,d^2\xi_\st}{(2\pi)^3}\ e^{i\,p{\cdot} \xi}
\ \langle P\vert \overline\psi_j(0)\,\psi_i(\xi)\vert P\rangle
\right|_{\xi{\cdot} n = 0} ,
\label{phi-TMD}
\end{equation}
and the collinear light-cone (LC) correlator
\begin{equation}
\Phi_{ij}(x) =
\left. \int \frac{d\xi{\cdot} P}{2\pi}\ e^{i\,p{\cdot} \xi}
\ \langle P\vert \overline\psi_j(0)\,\psi_i(\xi)\vert P\rangle
\right|_{\xi{\cdot} n = \xi_\st = 0\ \mbox{or}\ \xi^2 = 0} ,
\label{phi-x}
\end{equation}
or the local matrix element
\begin{equation}
\Phi_{ij} =
\left. \langle P\vert \overline\psi_j(0)\,\psi_i(\xi)\vert P\rangle
\right|_{\xi = 0} .
\end{equation}
The importance of integrating at least the light-cone (minus) component 
$p^- = p{\cdot} P$ is that the nonlocality is at equal (light-cone) time, 
i.e.\ time-ordering
is not relevant anymore for TMD or collinear PDFs~\cite{Diehl:1998sm}. 
For local matrix elements
one can calculate the anomalous dimensions, which show up as the Mellin 
moments of the splitting functions that govern the scaling behavior 
of the collinear correlator $\Phi(x)$. We note that the collinear 
correlator is not simply
an integrated TMD. The dependence on upper limit $\Phi(x;Q^2)
= \int^{Q}d^2p_\st\ \Phi(x,p_\st)$ is governed by the anomalous
dimensions (splitting functions). One has a $\alpha_s/p_\st^2$
behavior of TMDs that is calculable using collinear TMDs and which 
matches to the intrinsic 
non perturbative $p_\st$-behavior~\cite{Collins:1984kg}.
We note that in operator product expansion language, the collinear
correlators involve operators of definite twist, while TMD correlators
involve operators of various twist.

In order to determine the importance of a particular correlator in a 
hard process, one can do a dimensional analysis to find out when they 
contribute in an expansion in the inverse hard scale. Dominant are 
the ones with lowest
canonical dimension obtained by maximizing contractions with $n$, 
for instance for quark or gluon fields the minimal canonical dimensions 
dim[$\overline\psi(0)\rlap{/}n\,\psi(\xi)$] = 
dim[$F^{n\alpha}(0)\,F^{n\beta}(\xi)$] = 2, while an example 
for a multi-parton combination gives 
dim[$\overline\psi(0)\rlap{/}n\,A_\st^\alpha(\eta)\,\psi(\xi)$] = 3. 
Equivalently, one can maximize the number of $P$'s in the 
parametrization
of $\Phi_{ij}$. Of course one immediately sees that any number of
collinear $n{\cdot} A(\eta) = A^n(\eta)$ fields doesn't matter. Furthermore
one must take care of color gauge-invariance, for instance when dealing 
with the gluon fields and one must include derivatives in color 
gauge-invariant combinations. With dimension zero there is 
$iD^n = i\partial^n + gA^n$ and
with dimension one there is $iD_\st^\alpha = i\partial_\st^\alpha
+gA_\st^\alpha$.
The color gauge-invariant expressions for quark and gluon distribution 
functions actually include gauge link operators,
\begin{equation}
U_{[0,\xi]} 
= {\cal P}\exp\left(-i\int_0^{\xi} d\zeta_\mu A^\mu(\zeta)\right)
\end{equation}
connecting the nonlocal fields,
\begin{eqnarray}
& &\Phi_{q\,ij}^{[U]}(x,p_\st;n) =
\left. \int \frac{d\xi{\cdot} P\,d^2\xi_\st}{(2\pi)^3}\ e^{i\,p{\cdot} \xi}
\ \langle P\vert \overline\psi_j(0)\,U_{[0,\xi]}\,\psi_i(\xi)\vert P\rangle
\right|_{LF} ,
\\
& &\Phi_g^{[U,U^\prime]\,\mu\nu}(x,p_\st)
= \left. {\int}\frac{d(\xi{\cdot}P)\,d^2\xi_\st}{(2\pi)^3}\ e^{ip{\cdot}\xi}
\tr\,\langle P{,}S|\,F^{n\mu}(0)\,
U_{[0,\xi]}^{\phantom{\prime}}\,
F^{n\nu}(\xi)\,U_{[\xi,0]}^\prime\,
|P{,}S\rangle\right|_{LF} .
\end{eqnarray}
For transverse separations, the gauge links involve gauge links
running along the minus direction to $\pm \infty$ (dimensionally
preferred), which are closed with one or more transverse pieces 
at light-cone infinity~\cite{Belitsky:2002sm,Boer:2003cm}.
The two simplest possibilities are $U^{[\pm]}$ =
$U^n_{[0,\pm\infty]}\,U^T_{[0_\st,\xi_\st]}
\,U^n_{[\pm\infty,\xi]}$,
leading to gauge link dependent quark
TMDs $\Phi_q^{[\pm]}(x,p_\st)$. 
Which correlator is relevant in which process is a matter of
doing the (diagrammatic) calculation and 
resummation~\cite{Bacchetta:2005rm,Bomhof:2006ra}.
For gluons, the correlator involves 
color gauge-invariant traces of field operators $F^{n\alpha}$, which 
are written in the color-triplet representation, 
requiring the inclusion of \emph{two} gauge links $U_{[0,\xi]}$ and 
$U_{[\xi,0]}^\prime$. Thus even with the simplest links, one has 
already four gluon TMDs $\Phi_g^{[\pm,\pm]}(x,p_\st)$.

\section{Parametrization of TMDs}

In principle quark and gluon TMDs including a gauge link need to be 
parametrized with a set of PDFs 
(or in the case of fragmentation PFFs), which just as the
correlator depend on the gauge link $U$. For
quarks~\cite{Mulders:1995dh,Bacchetta:2006tn} one finds the
following set
of functions depending on $x$ and $p_\st^2$,
\begin{eqnarray}
&&\Phi^{[U]}(x,p_{\st};n) = \bigg\{
f^{[U]}_{1}(x,p_\st^2)
-f_{1T}^{\perp[U]}(x,p_\st^2)\,
\frac{\epsilon_{\st}^{p_{\st}S_{\st}}}{M}
+g^{[U]}_{1s}(x,p_\st)\gamma_{5}
\label{e:par}
\\&&\mbox{}\qquad
+h^{[U]}_{1T}(x,p_\st^2)\,\gamma_5\,\slashed{S}_{\st}
+h_{1s}^{\perp [U]}(x,p_\st)\,\frac{\gamma_5\,\slashed{p}_{\st}}{M}
+ih_{1}^{\perp [U]}(x,p_\st^2)\,\frac{\slashed{p}_{\st}}{M}
\bigg\}\frac{\slashed{P}}{2},
\nonumber
\end{eqnarray}
with the spin vector parametrized as 
$S^\mu = S_{\sL}P^\mu + S^\mu_{\st} + M^2\,S_{\sL}n^\mu$ 
and shorthand notations for $g^{[U]}_{1s}$ and $h_{1s}^{\perp [U]}$,
\begin{equation}
g^{[U]}_{1s}(x,p_\st)=S_{\sL} g^{[U]}_{1L}(x,p_{\st}^2)
-\frac{p_{\st}{\cdot} S_{\st}}{M}g^{[U]}_{1T}(x,p_{\st}^2).
\end{equation}
For quarks, these include not only the 
functions that survive upon $p_\st$-integration, $f_1^q(x) = q(x)$, 
$g_1^q(x) = \Delta q(x)$ and $h_1^q(x) = \delta q(x)$, which are the 
well-known collinear spin-spin densities 
(involving quark and
nucleon spin) but also momentum-spin densities such as the Sivers 
function
$f_{1T}^{\perp q}(x,p_\st^2)$ (unpolarized quarks in a transversely 
polarized nucleon) 
and spin-spin-momentum densities such as $g_{1T}(x,p_\st^2)$ 
(longitudinally polarized quarks in a transversely polarized nucleon). 
The parametrization for gluons,
following the naming convention in Ref.~\cite{Meissner:2007rx}, 
reads
\begin{eqnarray}
&&2x\,\Gamma^{\mu\nu [U]}(x{,}p_\st) = 
-g_T^{\mu\nu}\,f_1^{g [U]}(x{,}p_\st^2)
+g_T^{\mu\nu}\frac{\epsilon_T^{p_TS_T}}{M}
\,f_{1T}^{\perp g[U]}(x{,}p_\st^2)
\label{GluonCorr}
\\&&\mbox{}\qquad
+i\epsilon_T^{\mu\nu}\;g_{1s}^{g [U]}(x{,}p_\st)
+\bigg(\frac{p_T^\mu p_T^\nu}{M^2}\,
{-}\,g_T^{\mu\nu}\frac{p_\st^2}{2M^2}\bigg)\;h_1^{\perp g [U]}(x{,}p_\st^2)
\nonumber\\ &&\mbox{}\qquad
-\frac{\epsilon_T^{p_T\{\mu}p_T^{\nu\}}}{2M^2}\;
h_{1s}^{\perp g [U]}(x{,}p_\st)
-\frac{\epsilon_T^{p_T\{\mu}S_T^{\nu\}}
{+}\epsilon_T^{S_T\{\mu}p_T^{\nu\}}}{4M}\;
h_{1T}^{g[U]}(x{,}p_\st^2).
\nonumber
\end{eqnarray}
For TMD correlators time-reversal does not provide constraints
as future- and past-pointing is interchanged. Depending on the
behavior of the Dirac structure, one can distinguish
T-even and T-odd functions satisfying $f^{[U]} = \pm f^{[U^t]}$, where
$U^t$ is the time-reversed link of $U$. In the quark-quark
correlator $f_{1T}^{\perp q}$ and $h_1^{\perp q}$ are T-odd, in the
gluon-gluon correlator $f_{1T}^{\perp g}$, $h_{1T}^{g}$, 
$h_{1L}^{\perp g}$ and $h_{1T}^{\perp g}$ are T-odd. 
Since time reversal is a good symmetry of QCD, the appearance of T-even
or T-odd functions in the parametrization of the correlators is linked
to specific observables with this same character. In particular single
spin asymmetries are T-odd observables. 

\section{Moments of TMDs}

In order to study the gauge link dependence, it is convenient to construct moments of TMDs. The procedure of moments is well-known
for the moments of collinear functions. For $\Phi(x)$ in
Eq.~\ref{phi-x} one constructs moments
\begin{eqnarray}
x^{N}\Phi(x) & = & 
\left. \int \frac{d\xi{\cdot} P}{2\pi}\ e^{i\,p{\cdot} \xi}
\ \langle P\vert \overline\psi(0)\,(i\partial^n)^{N}
\,U^n_{[0,\xi]}\,\psi(\xi)\vert P\rangle
\right|_{LC} 
\\ & = & 
\left. \int \frac{d\xi{\cdot} P}{2\pi}\ e^{i\,p{\cdot} \xi}
\ \langle P\vert \overline\psi(0)\,U^n_{[0,\xi]}\,(iD^n)^{N}
\,\psi(\xi)\vert P\rangle
\right|_{LC}.
\nonumber
\end{eqnarray}
Integrating over $x$ one finds the connection of the Mellin moments of 
PDFs
with local matrix elements having specific anomalous dimensions,
which via an inverse Mellin transform define the splitting functions.
In the same way one can consider transverse moments~\cite{Boer:2003cm} 
starting with the light-front TMD in Eq.~\ref{phi-TMD},
\begin{eqnarray}
&&
p_\st^\alpha\,\Phi^{[\pm]}(x,p_\st;n) = 
\\&&
\mbox{}\qquad \left. \int \frac{d\xi{\cdot} P\,d^2\xi_\st}{(2\pi)^3}\ e^{i\,p{\cdot} \xi}
\ \langle P\vert \overline\psi(0)\,U^n_{[0,\pm\infty]}\,U^T_{[0_\st,\xi_\st]}
\,iD_\st^\alpha(\pm\infty)
\,U^n_{[\pm\infty,\xi]}\psi(\xi)\vert P\rangle
\right|_{LF} .
\nonumber
\end{eqnarray}
Integrating over $p_\st$ gives the lowest transverse moment. This 
moment involves 
twist-3 (or higher) collinear multi-parton correlators, in particular 
light-cone quark-quark-gluon correlator 
$\Phi^{n\alpha}_{F}(x-x_1,x_1|x)$
starting from Eq.~\ref{quarkgluonquark}, 
and the similarly defined correlator
$\Phi_D^\alpha(x-x_1,x_1|x)$. 
The particular combinations that are needed in the moments are 
\begin{eqnarray}
\widetilde\Phi_\partial^\alpha(x) 
& = & \Phi_D^\alpha(x) - \Phi_A^\alpha(x)
\\ & = &
\int dx_1\,\Phi_D^\alpha(x-x_1,x_1 | x)
-\int dx_1\ PV\frac{1}{x_1}\,\Phi_F^{n\alpha}(x-x_1,x_1 | x),
\nonumber
\\
\Phi_G^\alpha(x) &=& \pi\,\Phi_F^{n\alpha}(x,0 | x).
\end{eqnarray}
The latter is referred to as a gluonic pole or ETQS-matrix 
element~\cite{Efremov:1984ip,Qiu:1991pp,Boer:1997bw,Boer:2003cm}. 
The collinear correlators have trivial gauge links bridging 
the light-like
separation and thus they have no gauge link
dependence. The functions are T-even or T-odd under time
reversal. While the collinear function $\Phi(x)$ is T-even,
the functions contributing to the first moment involve the
T-even correlator $\widetilde\Phi_\partial^\alpha(x)$ and the 
T-odd correlator $\Phi_G^\alpha(x)$. For the higher moments
one finds that the relevant correlators involve similar
multi-partonic correlators, e.g.~for the double weighting
one needs the T-even correlator $\widetilde\Phi_{\partial\partial}^{\alpha\beta}(x)$,
the symmetrized T-odd correlator 
$\widetilde\Phi_{\{\partial G\}}^{\alpha\beta}(x)$ and the 
T-even double gluonic pole
correlator $\Phi_{GG}^{\alpha\beta}(x)$.
In terms of these functions we get for the moments
$\Phi_{\partial\ldots\partial}^{\alpha_1\ldots\alpha_n [U]}$
$\equiv$ $\int d^2p_\st\ p_\st^{\alpha_1}\ldots p_\st^{\alpha_n}
\,\Phi^{[U]}(x,p_\st)$
\begin{eqnarray}
&&
\Phi^{[U]}(x) = \Phi(x),
\quad
\Phi_\partial^{[U]\alpha}(x)
= \widetilde\Phi_{\partial}^{\alpha}(x) 
+ C_G^{[U]}\,\Phi_{G}^{\alpha}(x),
\\&&
\Phi_{\partial\partial}^{[U]\alpha\beta}(x)
= \widetilde\Phi_{\partial\partial}^{\alpha\beta}(x) 
+ C_G^{[U]}\,\Phi_{\{\partial G\}}^{\alpha\beta}(x)
+ \sum_{c=1}^2 C_{GG,c}^{[U]}\,\Phi_{GG,c}^{\alpha\beta}(x),
\end{eqnarray}
etc. The gauge link dependence is contained in calculable 
gluonic pole factors $C_G^{[U]}$, etc. For instance the 
factors $C_G^{[\pm]} = \pm 1$. An additional summation is
needed if there are multiple color arrangements possible
for the fields, e.g.\ the summation over $c$ in the double
gluonic pole contribution of the quark-quark correlator is
needed because one can have color structures 
$\tr(GG\psi\overline\psi)$ ($c = 1$) or 
$\tr(\psi\overline\psi)\tr(GG)/N_c$ ($c=2$). For the simplest
gauge links in quark correlators one has $C_{GG,1}^{[\pm]} = 1$
and $C_{GG,2}^{[\pm]} = 0$, but if Wilson loops appear in $U$
the latter coefficient is nonzero.

\section{Universal TMDs of definite rank}

The parametrization of quark-quark correlators in Eq.~\ref{e:par} 
shows that all weightings beyond double $p_\st$-weighting are
zero. From it we deduce for a nucleon~\cite{Buffing:2012sz}
\begin{eqnarray}
\Phi^{[U]}(x,p_\st) &\ =\ &
\Phi(x,p_\st^2) 
+\frac{p_{\st i}}{M}\,\widetilde\Phi_\partial^{i}(x,p_\st^2)
+\frac{p_{\st ij}}{M^2}
\,\widetilde\Phi_{\partial\partial}^{ij}(x,p_\st^2)
\label{eq:TMDstructure}
\\ &&
\mbox{} + C_{G}^{[U]}\left\lgroup\frac{p_{\st i}}{M}
\,\Phi_{G}^{i}(x,p_\st^2)
+ \frac{p_{\st ij}}{M^2}
\,\widetilde\Phi_{\{\partial G\}}^{\,ij}(x,p_\st^2)\right\rgroup
\nonumber \\ &&
\mbox{} + \sum_{c=1}^2 C_{GG,c}^{[U]}\,\frac{p_{\st ij}}{M^2}
\,\Phi_{GG,c}^{ij}(x,p_\st^2),
\nonumber
\end{eqnarray}
where $p_\st^{ij}$ is the second rank traceless tensor
$p_\st^{ij}$ = $p_\st^i p_\st^j - \frac{1}{2}\,p_\st^2\,g_\st^{ij}$.
We refer to the correlators in Eq.~\ref{eq:TMDstructure} as 
TMD correlators of definite rank. Each of these definite rank 
correlators is parametrized in terms of universal TMD PDFs. Using
the T-behavior of the functions and their tensorial structure
the identification is straightforward, with for instance
$f_{1T}^{\perp}$ and $h_1^\perp$ appearing in $\Phi_G$. 
Furthermore, one obtains
\begin{eqnarray}
&& f_1^{[U]}(x,p_\st^2) = f_1(x,p_\st^2),
\quad g_{1T}^{[U]}(x,p_\st^2) = g_{1T}(x,p_\st^2), \quad \ldots ,
\\&&
f_{1T}^{\perp [U]}(x,p_\st^2) = C_G^{[U]}\,f_{1T}^\perp(x,p_\st^2),
\quad h_1^{\perp [U]}(x,p_\st^2) = C_G^{[U]}\,h_1^\perp(x,p_\st^2),
\\&&
h_{1T}^{\perp [U]}(x,p_\st^2) = h_{1T}^{\perp (A)}(x,p_\st^2)
+ C_{GG,1}^{[U]}\,h_{1T}^{\perp (B1)}(x,p_\st^2)
+ C_{GG,2}^{[U]}\,h_{1T}^{\perp (B2)}(x,p_\st^2).
\end{eqnarray}
Functions like $f_1$ and $g_{1T}$ are universal. The T-odd rank-1 TMD 
PDFs like $h_1^\perp$ also are universal but appear in cross sections
with calculable process dependent gluonic pole factors. At rank-2,
however, one has three universal T-even pretzelocity functions 
$h_{1T}^{\perp \ldots}$ appearing in different, but 
calculable,combinations in hard cross sections.
For a spin 1/2 target the rank-2 quark-quark correlator 
$\Phi_{\{\partial G\}}$ = 0, but such a rank-2 correlator is
relevant for a spin one hadron~\cite{Buffing:2012sz} or 
for gluon-gluon TMD correlators~\cite{BMM2}.
The analogous treatment for fragmentation functions is simpler.
In that case there is no process dependence~\cite{Collins:2004nx}.
Phrased in terms of operators, the gluonic pole matrix elements 
vanish in that 
case~\cite{Meissner:2008yf,Gamberg:2008yt,Gamberg:2010gp}. 
Nevertheless, there exist T-odd fragmentation functions, but 
their QCD operator structure is T-even, 
similar to the structure of $\widetilde\Phi_\partial^\alpha$. 

The universal $p_\st^2$-dependent TMD functions and correlators 
show up in azimuthal asymmetries, with the azimuthal dependence
contained in the tensors in Eq.~\ref{eq:TMDstructure}. For a given 
rank $m$ they just are $\vert p_\st\vert^m\,e^{\pm im\varphi}$. 
For the study of the $p_\st^2$-dependence, it is convenient 
to use Bessel transforms~\cite{Boer:2011xd} 
of the $(-p_\st^2/2M^2)$-moments,
\begin{equation}
f^{(m/2)}_{\ldots}(x,p_\st^2) =
\left(\frac{-p_\st^2}{2M^2}\right)^{m/2}\,f_{\ldots}(x,p_\st^2)
= \int_0^\infty db\ \sqrt{\vert p_\st\vert b}\,J_m(\vert p_\st\vert b)
\,f^{(m/2)}_{\ldots}(x,b).
\end{equation}
For quarks one gets multiple color possibilities at rank two,
leading to three universal pretzelocities. For gluons~\cite{BMM2} 
there are more situations in which multiple color configurations
show up, among them in the study of the gluon Boer-Mulders function
$h_1^{\perp g}(x,p_\st)$ corresponding to linear gluon polarization 
in an unpolarized nucleon.

\acknowledgments
This research is part of the research program of the Foundation for Fundamental Research of Matter (FOM) and the Netherlands Organization
for Scientific Research (NWO), and the FP7 EU-programme HadronPhysics3 (contract no 283286).


\begin{thebibliography}{00}

\bibitem{Diehl:1998sm} M. Diehl and T. Gousset, {\it Phys. Lett. B\/} 
{\bf 428}, 359 (1998).

\bibitem{Collins:1984kg} J.C. Collins and D.E. Soper 
and G.F. Sterman, {\it Nucl. Phys. B\/} {\bf 250}, 199 (1985).

\bibitem{Belitsky:2002sm} A.V. Belitsky, X. Ji and F. Yuan, 
{\it Nucl. Phys. B\/} {\bf 656}, 165 (2003).

\bibitem{Boer:2003cm} D. Boer, P.J. Mulders and F. Pijlman, 
{\it Nucl. Phys. B\/} {\bf 667}, 201 (2003).

\bibitem{Bacchetta:2005rm} A. Bacchetta, C.J. Bomhof, P.J. Mulders 
and F. Pijlman,
{\it Phys. Rev. D} {\bf 72}, 034030 (2005). 

\bibitem{Bomhof:2006ra} C.J. Bomhof and P.J. Mulders, {\it JHEP} 
{\bf 0702}, 029 (2007).

\bibitem{Mulders:1995dh} P.J. Mulders and R.D. Tangerman, 
{\it Nucl. Phys. B\/} {\bf 461}, 197 (1996)
[Erratum-ibid. {\it B\/} {\bf 484}, 538 (1997)].

\bibitem{Bacchetta:2006tn} A. Bacchetta et al., 
{\it JHEP} {\bf 0702}, 093 (2007).

\bibitem{Meissner:2007rx} S. Meissner, A. Metz and K. Goeke, 
{\it Phys. Rev. D\/} {\bf 76}, 034002 (2007).

\bibitem{Efremov:1984ip} A.V. Efremov and O.V. Teryaev, 
{\it Phys. Lett. B\/} {\bf 150}, 383 (1985).

\bibitem{Qiu:1991pp} J-W. Qiu and G.F. Sterman, {\it Phys. Rev. Lett.} 
{\bf 67}, 2264 (1991).

\bibitem{Boer:1997bw} D. Boer, P.J. Mulders and O.V. Teryaev,
{\it Phys. Rev.} {\bf D57}, 3057 (1998).

\bibitem{Buffing:2012sz} M.G.A. Buffing, A. Mukherjee 
and P.J. Mulders, {\it Phys. Rev. D\/} 
{\bf 86}, 074030 (2012).

\bibitem{BMM2} M.G.A. Buffing, A. Mukherjee and P.J. Mulders, {\it 
Generalized Universality of Definite Rank Gluon Transverse
Momentum Dependent Correlators}, in preparation.

\bibitem{Collins:2004nx} J.C. Collins and A. Metz, 
{\it Phys. Rev. Lett.} {\bf
93}, 252001 (2004).

\bibitem{Meissner:2008yf} S. Meissner and A. Metz, 
{\it Phys. Rev. Lett.} {\bf
102}, 172003 (2009).

\bibitem{Gamberg:2008yt} L.P. Gamberg, A. Mukherjee and P.J. Mulders,
{\it Phys. Rev. D\/} {\bf 77}, 114026 (2008). 

\bibitem{Gamberg:2010gp} L.P. Gamberg, A. Mukherjee and P.J. Mulders,
{\it Phys. Rev. D\/} {\bf 83}, 071503 (2011).

\bibitem{Boer:2011xd} D. Boer, L. Gamberg, B. Musch and A. Prokudin,
{\it JHEP} {\bf 10} 021 (2011).

\end{thebibliography}
\end{document}